\documentclass[12pt]{article}
\renewcommand{\today}{September 1996}

\begin{document}
\sloppy
\begin{titlepage}
\null
\vspace{5mm}
\begin{flushright}
\begin{tabular}{l}
DFTT 63/96\\
hep-ph/9610314\\
\today
\end{tabular}
\end{flushright}
\vfill
\begin{center}
\Large
\textbf{POSSIBLE TESTS OF
THE ``STRANGENESS'' OF THE NUCLEON}\footnote
{\normalsize
Talk presented by S.M. Bilenky at the
$13^{\mathrm{th}}$
\textit{International seminar on high-energy physics problems}
(ISHEPP 96):
\textit{relativistic nuclear physics and quantum chromodynamics},
Dubna, Russia, September 2-7, 1996.}
\\[5mm]
\normalsize
W.M. Alberico$^{\mathrm{a}}$,
S.M. Bilenky$^{\mathrm{a,b}}$
and
C. Giunti$^{\mathrm{a}}$
\\[3mm]
(a) Dipartimento di Fisica Teorica,
Universit\`a di Torino and INFN,
Sezione di Torino,\\
Via P. Giuria 1, I-10125 Torino, Italy
\\
(b) Joint Institute for Nuclear Research, Dubna, Russia
\end{center}
\vfill
\begin{center}
\textbf{Abstract}
\\[3mm]
\begin{minipage}{0.8\textwidth}
The NC and CC scattering
of neutrino(antineutrino)
on protons and nuclei
with isotopic spin equal to zero is considered.
It is shown that the measurement of
the neutrino-antineutrino asymmetry
in these processes could
allow to obtain
model-independent
information about the contribution of
the axial strange current
to the cross sections of the NC processes.
\end{minipage}
\end{center}
\vfill
\null
\end{titlepage}

The problem of the ``strangeness'' of nucleon
is a very
important theoretical and experimental issue
(see, for example, Ref.\cite{Anselmino}).
The data obtained in the recent CERN \cite{CERN}
and SLAC \cite{SLAC}
experiments on deep inelastic scattering
of polarized leptons on polarized nucleons  
are in agreement with the data of EMC collaboration
\cite{EMC}.
From the 
analysis of these data it follows that
the one-nucleon matrix elements of the axial
strange current is relatively 
large and comparable with the matrix elements of
axial $u$ and $d$ currents.
In a recent analysis of the data
\cite{Ellis-Karliner}
it was found that
\begin{equation}
\begin{array}{l} \displaystyle
\Delta u
=
0.82 \pm 0.03
\;,
\\ \displaystyle
\Delta d
=
- 0.44 \pm 0.03
\;,
\\ \displaystyle
\Delta s
=
- 0.11 \pm 0.03
\;,
\end{array}
\label{01}
\end{equation}
where the constants
$ \Delta q $
(with $q=u,d,s$)
are determined by
\begin{equation}
\left\langle
p \, , \, r
\left|
\,
\bar{q} \, \gamma^{\alpha} \, \gamma_{5} \, q
\,
\right|
p \, , \, r
\right\rangle
=
2 M
\,
r
\,
s^{\alpha}
\,
\Delta q
\;.
\label{02}
\end{equation}
Here
$M$ is the nucleon mass
and
$
\left|
p \, , \, r
\right\rangle
$
is the state vector of a nucleon
with momentum $p$ and projection of the spin
on the direction
$s^{\alpha}$
equal to $r$.

As it is well known,
in order to determine the  
constants
$ \Delta q $
from the data on polarized deep inelastic
lepton-nucleon scattering
it is necessary extrapolate
the polarized structure function $g_1$ to
the point $x=0$.
Usually,
a Regge behaviour of the function $g_1$ is assumed.
The value of the constant $F/D$ that is also used
in order to determine the constants
$ \Delta q $
is obtained from
the data on the leptonic decays of hyperons
under the assumption
that the effects of violation of SU(3) can be neglected.
Both these assumption require further investigation
(see, for example, Ref.\cite{Cheng96}).

Alternative methods
that could allow
to determine the matrix elements
of the strange currents are
obviously necessary.
It is well known
(see Ref.\cite{Kaplan-Manohar})
that NC-induced processes
could be important sources of information 
on the matrix elements of the axial and vector 
strange currents.

The neutral current of the Standard Model
in the $u$, $d$ and $s$ approximation
has the form
\begin{equation}
j^Z_{\alpha}
=
v^{3}_{\alpha}
-
2 \sin^2\theta_{W}
\,
j^{\mathrm{em}}_{\alpha}
+
a^{3}_{\alpha}
-
\frac{1}{2}
\,
v^{S}_{\alpha}
-
\frac{1}{2}
\,
a^{S}_{\alpha}
\;,
\label{03}
\end{equation}
where
$v^{3}_{\alpha}$
and
$a^{3}_{\alpha}$
are the third
components of the isovector vector and axial currents,
$j^{\mathrm{em}}_{\alpha}$
is the electromagnetic current and
\begin{equation}
v^{S}_{\alpha}
=
\bar{s} \gamma_{\alpha} s
\;,
\qquad
a^{S}_{\alpha}
=
\bar{s} \gamma_{\alpha} \gamma_{5} s
\label{04}
\end{equation}
are the vector and axial strange currents.
 
Using Eq.(\ref{03}) we can connect
the matrix elements
of the strange currents with the quantities that
can be measured with the investigation of NC, CC and
electromagnetic processes.
This will be our main strategy.

There are
two types of NC-induced effects:
\begin{enumerate}
\item
P-odd effects in lepton-nucleon scattering;
\item
NC neutrino-nucleon(nucleus) scattering.
\end{enumerate}
P-odd effects 
in lepton-nucleon scattering are due to
the interference of diagrams
with $\gamma$ and $Z$ exchange.
It is easy to see from
general symmetry arguments that the contribution  
of the vector (axial) strange currents 
to the P-odd asymmetry is proportional
to
$ g_V = - 1/2 + 2\sin^2\theta_W $
($ g_A = - 1/2 $).
Using the latest value of the parameter
$\sin^2\theta_W$
($ \sin^2 \theta_{W} = 0.226 \pm 0.004 $
\cite{PDG96})
we obtain 
$ g_V \simeq - 0.05 $.
Thus,
the investigation of
P-odd effects in lepton-nucleon scattering could
allow
to obtain information only about
the matrix elements of the strange \emph{vector} current.
On the other hand, 
in the cross sections of neutrino
processes there is no any apriori
suppression of the contributions
of both axial and vector
strange currents.

In Ref.\cite{ABGM96}
we have considered in detail the possibilities
to extract information about the strange vector and axial
nucleon form factors from the investigation of
the elastic
neutrino(antineutrino)-nucleon scattering
processes
\begin{equation}
\nu_{\mu}
\,
(\bar\nu_{\mu})
+
p
\to
\nu_{\mu}
\,
(\bar\nu_{\mu})
+
p
\;.
\label{05}
\end{equation}
We have shown that
the strange magnetic
$G_M^S(Q^2)$
and
strange axial
$F_A^S(Q^2)$
form factors are connected
with asymmetry
\begin{equation}
A(Q^2)
=
\frac
{
\left(
\mathrm{d}\sigma / \mathrm{d}Q^2
\right)^{\mathrm{NC}}_{\nu{p}}
-
\left(
\mathrm{d}\sigma / \mathrm{d}Q^2
\right)^{\mathrm{NC}}_{\bar\nu{p}}
}
{
\left(
\mathrm{d}\sigma / \mathrm{d}Q^2
\right)^{\mathrm{CC}}_{\nu{n}}
-
\left(
\mathrm{d}\sigma / \mathrm{d}Q^2
\right)^{\mathrm{CC}}_{\bar\nu{p}}
}
\label{06}
\end{equation}
by the relation
\begin{equation}
4 \, | V_{ud} |^2 \,
A(Q^2)
=
1 - 2 \sin^2\theta_W
\,
\frac{G_M^{p}(Q^2)}{G_M^3(Q^2)}
-
\left(
1 - 2 \sin^2\theta_W
\,
\frac{G_M^{p}(Q^2)}{G_M^3(Q^2)}
\right)
\frac{F^s_A(Q^2)}{F_A(Q^2)}
-
\frac{1}{2}
\,
\frac{G_M^s(Q^2)}{G_M^3(Q^2)}
\;.
\label{07}
\end{equation}
Here $V_{ud}$ is the $ud$ element of
the CKM matrix,
$G_M^3(Q^2)=(G_M^{p}(Q^2)-G_M^{n}(Q^2))/2$
is the isovector magnetic form factor of the nucleon,
$G_M^{p(n)}(Q^2)$
is the magnetic form factor of the proton (neutron),
$F_A(Q^2)$
is the CC axial
form factor,
$F^s_A(Q^2)$
and
$G_M^s(Q^2)$
are axial and magnetic strange
form factors.
In Eq.(\ref{06})
the quantities
$
\left(
\mathrm{d}\sigma / \mathrm{d}Q^2
\right)^{\mathrm{NC}}_{\nu{p}}
$
and
$
\left(
\mathrm{d}\sigma / \mathrm{d}Q^2
\right)^{\mathrm{NC}}_{\bar\nu{p}}
$
are the differential cross sections
of the NC elastic processes (\ref{05}),
and
$
\left(
\mathrm{d}\sigma / \mathrm{d}Q^2
\right)^{\mathrm{CC}}_{\nu{n}}
$
and
$
\left(
\mathrm{d}\sigma / \mathrm{d}Q^2
\right)^{\mathrm{CC}}_{\bar\nu{p}}
$
are the differential cross sections
of the CC quasi-elastic processes
\begin{eqnarray}
&&
\nu_\mu + n \to \mu^{-} + p
\;,
\label{09}
\\
&&
\bar\nu_\mu + p \to \mu^{+} + n
\;.
\label{10}
\end{eqnarray}
The relation (\ref{07}) depends only
on the magnetic form factors of
the nucleon,
which are known 
from the experimental data with 
better accuracy than
the electric form factors.
Moreover,
in this relation 
enters the ratio
of the magnetic form factors of
the neutron and proton,
that is approximately equal to the ratio
of the corresponding magnetic
moments (scaling law).
A measurement of the asymmetry
$A(Q^2)$
could allow to obtain
information about the strange axial and vector form factors
of the nucleon.

We will consider now
the \emph{inclusive} neutrino processes
\begin{equation}
\nu_{\mu}
\,
(\bar\nu_{\mu})
+
A
\to
\nu_{\mu}
\,
(\bar\nu_{\mu})
+
X
\;,
\label{11}
\end{equation}
where $A$
is a nucleus with isotopic spin $T$ equal to zero
(as $d$, $^{4}$He, $^{12}$C, etc.).

The cross sections of the processes
(\ref{11})
have the following general form
\begin{equation}
\left(
\frac
{ \mathrm{d}\sigma }
{ \mathrm{d}Q^2 \mathrm{d}\nu }
\right)^{\mathrm{NC}}_{\nu(\bar\nu)}
=
\frac{G_F^2}{4\pi}
\left(
\frac{M}{p{\cdot}k}
\right)^2
\left(
L^{\alpha\beta}(k,k')
\mp
L_{5}^{\alpha\beta}(k,k')
\right)
W^{\mathrm{NC}}_{\alpha\beta}(p,q)
\;.
\label{12}
\end{equation}
where
$M$ is the mass of the nucleus, 
$p$ is the nucleus momentum,
$k$ and $k'$ are the  momenta of the initial and
final neutrinos (antineutrinos),
$ q = k-k' $,
$ Q^2 = - q^2 $,
\begin{eqnarray}
L^{\alpha\beta}(k,k')
\null & = & \null
k^{\alpha} \, {k'}^{\beta}
-
g^{\alpha\beta} \, k \cdot k'
+
{k'}^{\alpha} \, k^{\beta}
\;,
\label{13}
\\
L_{5}^{\alpha\beta}(k,k')
\null & = & \null
i
\,
\epsilon^{\alpha\beta\rho\sigma}
\,
k_{\rho}
\,
k'_{\sigma}
\;,
\label{14}
\end{eqnarray}
and
the hadronic tensor
$W^{\mathrm{NC}}_{\alpha\beta}(p,q)$
is given by
\begin{equation}
W^{\mathrm{NC}}_{\alpha\beta}(p,q)
=
( 2\pi )^2
\,
\frac{ p^0 }{ M }
\int
\langle
p
|
J^{Z}_{\beta}(x)
\,
J^{Z}_{\alpha}(0)
|
p
\rangle
\,
\mathrm{e}^{ - i q \cdot x }
\,
\mathrm{d}^4x
\;.
\label{15}
\end{equation}

In order to separate the contribution of strange axial
current,  
let us consider the difference
of the cross sections of the processes
(\ref{11})
From Eq.(\ref{12})
we have
\begin{equation}
\left(
\frac
{ \mathrm{d}\sigma }
{ \mathrm{d}Q^2 \mathrm{d}\nu }
\right)^{\mathrm{NC}}_{\nu}
-
\left(
\frac
{ \mathrm{d}\sigma }
{ \mathrm{d}Q^2 \mathrm{d}\nu }
\right)^{\mathrm{NC}}_{\bar\nu}
=
-
\frac{G_F^2}{2\pi}
\left( \frac{ M }{ p \cdot k } \right)^2
L_{5}^{\alpha\beta}(k,k')
\,
W^{\mathrm{NC};I}_{\alpha\beta}(p,q)
\;,
\label{16}
\end{equation}
where
$W^{\mathrm{NC};I}_{\alpha\beta}(p,q)$
is the contribution of the interference
of the vector and axial currents to
$W^{\mathrm{NC}}_{\alpha\beta}(p,q)$.
In the case of
a target with isospin $T=0$,
the interference of the isovector and
isoscalar currents do not contribute to
$W^{\mathrm{NC}}_{\alpha\beta}(p,q)$
and we have
\begin{equation}
W^{\mathrm{NC};I}_{\alpha\beta}(p,q)
=
\left(
1
-
2 \sin^2 \theta_{W}
\right)
W^{V^{3};A^{3}}_{\alpha\beta}(p,q)
+
\sin^2 \theta_{W}
\,
W^{V^{0};A^{S}}_{\alpha\beta}(p,q)
\;,
\label{17}
\end{equation}
where
\begin{eqnarray}
W^{V^{3};A^{3}}_{\alpha\beta}(p,q)
\null & = & \null
( 2\pi )^2
\,
\frac{ p^0 }{ M }
\int
\langle
p
|
V^{3}_{\beta}(x)
\,
A^{3}_{\alpha}(0)
+
A^{3}_{\beta}(x)
\,
V^{3}_{\alpha}(0)
|
p
\rangle
\,
\mathrm{e}^{ - i q \cdot x }
\,
\mathrm{d}^4x
\;,
\label{18}
\\
W^{V^{0};A^{S}}_{\alpha\beta}(p,q)
\null & = & \null
( 2\pi )^2
\,
\frac{ p^0 }{ M }
\int
\langle
p
|
V^{0}_{\beta}(x)
\,
A^{S}_{\alpha}(0)
+
A^{S}_{\beta}(x)
\,
V^{0}_{\alpha}(0)
|
p
\rangle
\,
\mathrm{e}^{ - i q \cdot x }
\,
\mathrm{d}^4x
\;.
\label{19}
\end{eqnarray}
Let us notice that
in Eq.(\ref{17})
we neglected the term quadratic in the strange currents.

The first term in the Eq.(\ref{17}) is connected with
the corresponding pseudotensor that determines the
difference of the cross sections of inclusive CC processes
\begin{equation}
\nu_{\mu}
\,
(\bar\nu_{\mu})
+
A
\to
\mu^{-}
\,
(\mu^{+})
+
X
\;.
\label{20}
\end{equation}
In fact,
in the region of relatively small energies
(up to the threshold of charm production)
the main contribution to the cross sections of the processes
(\ref{20})
comes from the $ud$ part of the charged hadronic current:
\begin{equation}
j_{\alpha}^{+}
=
V_{ud}
\left(
v_{\alpha}^{1+i2}
+
a_{\alpha}^{1+i2}
\right)
\;,
\label{21}
\end{equation}
where
$
v_{\alpha}^{1+i2}
=
v_{\alpha}^{1}
+
i \, v_{\alpha}^{2}
$
and
$
a_{\alpha}^{1+i2}
=
a_{\alpha}^{1}
+
i \, a_{\alpha}^{2}
$.
For the difference of the cross sections
of the processes (\ref{20})
we have
\begin{equation}
\left(
\frac
{ \mathrm{d}\sigma }
{ \mathrm{d}Q^2 \mathrm{d}\nu }
\right)^{\mathrm{CC}}_{\nu}
-
\left(
\frac
{ \mathrm{d}\sigma }
{ \mathrm{d}Q^2 \mathrm{d}\nu }
\right)^{\mathrm{CC}}_{\bar\nu}
=
-
\frac{G_F^2}{2\pi}
\left( \frac{ M }{ p \cdot k } \right)^2
L_{5}^{\alpha\beta}(k,k')
\,
W^{\mathrm{CC};I}_{\alpha\beta}(p,q)
\,
|V_{ud}|^2
\;,
\label{22}
\end{equation}
where
$W^{\mathrm{CC};I}_{\alpha\beta}(p,q)$
is the contribution of the interference
of the hadronic vector and axial currents to
$W^{\mathrm{CC}}_{\alpha\beta}(p,q)$.

From the isotopic SU(2) invariance of the strong interactions
it follows that
\begin{equation}
W^{\mathrm{CC};I}_{\alpha\beta}(p,q)
=
2
\,
W^{V^{3};A^{3}}_{\alpha\beta}(p,q)
\;.
\label{23}
\end{equation}

For the neutrino--antineutrino asymmetry
\begin{equation}
A(Q^2,\nu)
=
\frac
{
\left(
\mathrm{d}\sigma / \mathrm{d}Q^2 \mathrm{d}\nu
\right)^{\mathrm{NC}}_{\nu}
-
\left(
\mathrm{d}\sigma / \mathrm{d}Q^2 \mathrm{d}\nu
\right)^{\mathrm{NC}}_{\bar\nu}
}
{
\left(
\mathrm{d}\sigma / \mathrm{d}Q^2 \mathrm{d}\nu
\right)^{\mathrm{CC}}_{\nu}
-
\left(
\mathrm{d}\sigma / \mathrm{d}Q^2 \mathrm{d}\nu
\right)^{\mathrm{CC}}_{\bar\nu}
}
\;,
\label{24}
\end{equation}
from Eqs.(\ref{16}), (\ref{17}), (\ref{22}) and (\ref{23}),
we obtain the following relation
\begin{equation}
A(Q^2,\nu)
\,
|V_{ud}|^2
=
\frac{1}{2}
\left(
1
-
2 \sin^2 \theta_{W}
\right)
+
\frac{1}{2}
\,
\sin^2 \theta_{W}
\,
\frac
{ L_{5}^{\alpha\beta} \, W^{V^{0};A^{S}}_{\alpha\beta} }
{ L_{5}^{\alpha\beta} \, W^{V^{3};A^{3}}_{\alpha\beta} }
\;.
\label{25}
\end{equation}
Thus,
a measurement of the asymmetry
$A(Q^2,\nu)$
could allow to determine the relative contribution
of the axial strange current to the cross sections
of the NC inclusive processes (\ref{11})
in a direct model independent way.

It was shown in Ref.\cite{Paschos-Wirbel}
that in the deep inelastic
region the last term of the relation (\ref{25}) is small.
The corresponding relation for the total cross sections
(the so-called Paschos-Wolfenstein relation
\cite{Paschos-Wolfenstein})
has been used in the deep inelastic region for
the determination of the value
of the parameter
$\sin^2 \theta_{W}$
from CC and NC neutrino and
antineutrino data.
We are proposing to use relation (\ref{25})
at intermediate neutrino energies, 
in which the parton approximation is not
applicable and nuclei behave
as bound states of nucleons. 
It was shown in Ref.\cite{Donnelly92} that
in this energy region
the contribution
of the isoscalar part of the electromagnetic current
to the longitudinal part of the
cross section of electron--carbon inclusive scattering
is comparable with the contribution
of the isovector part of the current.
If there is no dynamical suppression
of the contribution of the axial strange
current to the asymmetry
$A(Q^2,\nu)$
at intermediate
neutrino energies,
then,
in accordance with the data from
EMC and other experiments,
the last term in the relation (\ref{25})
can be significant.
A significant deviation of the measured asymmetry
$A(Q^2,\nu)$
from the known quantity
$
( 1 - 2 \sin^2 \theta_{W} ) / 2 |V_{ud}|^2
=
0.289 \pm 0.004
$
(derived from
$ \sin^2 \theta_{W} = 0.226 \pm 0.004 $
and
$ |V_{ud}| = 0.9736 \pm 0.0010 $
\cite{PDG96})
would be a clear signal in favor of the importance of the
contribution of the axial strange 
current to the hadronic matrix elements.

\bigskip

We would like
to express our deep gratitude to A. Molinari for the
useful discussions on the problem of the isoscalar current.


\begin{thebibliography}{99}

\bibitem{Anselmino}
M. Anselmino, A. Efremov and E. Leader,
Phys. Rep. {\bf 261}, 1 (1995).

\bibitem{CERN}
D. Adams et al.,
Phys. Lett. B {\bf 329}, 399 (1994).

\bibitem{SLAC}
K. Abe et al.,
Phys. Rev. Lett. {\bf 74}, 346 (1995).

\bibitem{EMC}
J. Ashman {\it et al.},
Phys. Lett. B {\bf 206}, 364 (1988);
Nucl. Phys B {\bf 328},1 (1989).

\bibitem{Ellis-Karliner}
J. Ellis and M. Karliner,
preprint
hep-ph/9601280.

\bibitem{Cheng96}
Hai-Yang Cheng,
preprint
hep-ph/9607254.

\bibitem{Kaplan-Manohar}
D.B. Kaplan and A. Manohar,
Nucl. Phys. B \textbf{310}, 527 (1988);
J. Ellis and M. Karliner,
Phys. Lett. B \textbf{213}, 73 (1988).

\bibitem{PDG96}
R.M. Barnett et al.,
Phys. Rev. D \textbf{54}, 1 (1996).

\bibitem{ABGM96}
W.M. Alberico, S.M. Bilenky, C. Giunti and C. Maieron,
Z. Phys. C \textbf{70}, 463 (1996).

\bibitem{Paschos-Wirbel}
E.A. Paschos and M. Wirbel,
Nucl. Phys. B \textbf{194} , 189 (1982).

\bibitem{Paschos-Wolfenstein}
E.A. Paschos and L. Wolfenstein,
Phys. Rev. D \textbf{7}, 91 (1973).

\bibitem{Donnelly92}
M.B. Barbaro et al.,
Nucl. Phys. A \textbf{598}, 503 (1996).

\end{thebibliography}
\end{document}